\documentclass[aps,prb,showpacs,twocolumn]{revtex4}

\usepackage{amssymb}
\usepackage{graphicx}
\usepackage{amsmath}
\usepackage{color}

\def\vo2{VO$_2$}
\def\t2g{$t_{2g}$}
\def\eg{$e_{g}$\;}

\def\egsi{$e_{g}^{\sigma}$\;}
\def\a1g{$a_{1g}$}

\def\m1{$M_1$}

\begin{document}

\title{Evidence for strong Coulomb correlations in metallic phase of vanadium dioxide}

\author{A.S.~Belozerov}
\affiliation{Institute of Metal Physics, Russian Academy of Sciences, 620041,
Ekaterinburg GSP-170, Russia}
\author{A.I.~Poteryaev}
\affiliation{Institute of Metal Physics, Russian Academy of Sciences, 620041,
Ekaterinburg GSP-170, Russia}
\author{V.I.~Anisimov}
\affiliation{Institute of Metal Physics, Russian Academy of Sciences, 620041,
Ekaterinburg GSP-170, Russia}

\date{\today}

\begin{abstract}
The influence of Coulomb correlation on magnetic and spectral properties
in metallic rutile phase of vanadium dioxide is studied by state of the art LDA+DMFT method.
Calculation results in strongly correlated metallic state
with an effective mass renormalization $m^*/m\approx2$.
Uniform  magnetic susceptibility shows
Curie-Weiss temperature dependence with effective magnetic moment, $p_{eff}^{theor} = 1.54 \mu_B$,
in a good agreement with the experimental value  $p_{eff}^{exp} = 1.53 \mu_B$
that is close to ideal value for V$^{4+}$ ion with the spin $S=1/2$,
$p_{eff} = 1.73 \mu_B$.
Calculated spectral function shows well developed Hubbard bands
observabale in the recent experimental photoemission spectra.
We conclude that \vo2 in rutile phase is strongly correlated metal
with local magnetic moments formed by vanadium $d$-electrons.
\end{abstract}


\maketitle

\begin{figure}
  \centering
  \includegraphics[clip=true, width=0.4\textwidth]{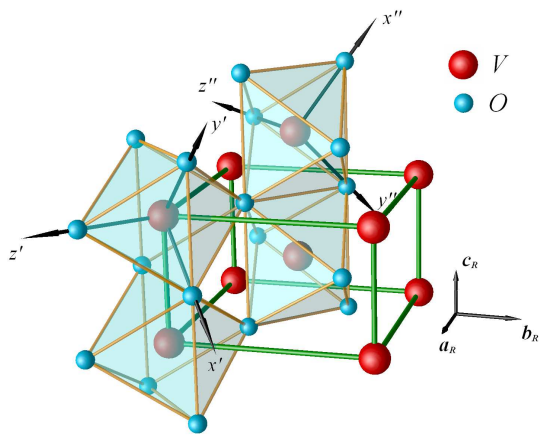}
  \caption{(Color online)  Crystal structure of rutile phase. Vanadium and oxygen atoms are 
           denoted by large (red) and small (cyan) balls.
           $(a_R,b_R,c_R)$ are rutile crystallographic axes and coincide with the Cartesian one.
           Local coordinate axes on the vanadium atoms are shown by (double) primed set.}
  \label{fig:structure_r}
\end{figure}

Transition metal compounds and particularly vanadium oxides attract a lot of interests
because of rich variety of structural modifications and properties~\cite{Imada1998}.
Vanadium dioxide is not an exception and is well known due to metal-insulator
transition at 340~K~\cite{Villeneuve1977Pouget1974}. The value of the energy gap ($\Delta_g$=0.6~eV)
in insulating phase and vicinity of the transition and room temperatures
allows one to use this material as ``wisdom`` windows and
in another termochromic applications~\cite{Chain1991}.
This compound has been extensively investigated experimentally and theoretically
over the half of century.
The phase diagram~\cite{Villeneuve1977Pouget1974} in the temperature-pressure (or chemical doping) coordinates
consists of the paramagnetic metal at high temperature and three insulating states
below $T_c$. Below transition temperature there are three insulating phases, called \m1, $T$
and $M_2$ at low, intermediate and large values of doping, respectively.
The \m1 and $M_2$ phases crystallize to monoclinic structure which can be obtained
from high-temperature rutile by distortion and doubling of the cell.
In the former structure equivalent vanadium atoms are paired and off-axis shifted
forming zigzags while in the later one there are two kinds of vanadium chains,
one of which is paired and another one forms zigzags without pairing.
In both phases the conductivity drops by factor of several orders in magnitude comparing with rutile phase
with the estimation of the energy gap about 0.6~eV~\cite{Berglund1969}.
Above the transition temperature \vo2 crystallizes to rutile structure
at all values of doping and can be regarded as two vanadium chains along $c_R$ axis
(see Fig.~\ref{fig:structure_r}).

Most of earlier theoretical studies for correlation effects in vanadium dioxide were concentrated
on the insulating low-temperature phases where band structure calculations have difficulties to reproduce
correct insulating type of electronic structure~\cite{Goodenough1971,Rice1994Wentzcovitch1994,Eyert2002}.
Calculation for rutile crystal structure resulted in metallic bands
in agreement with experiment, and hence, it was assumed that taking into account
Coulomb correlations is not needed here. However magnetic and spectral measurements
for \vo2 show that correlation strength could be essential
for metallic rutile phase~\cite{Okazaki2004,Okazaki2006}.
The magnetic experiments for metallic rutile phase show a quite large temperature dependent
magnetic susceptibility
that indicates  formation of local magnetic moment in the partially filled $d$-band.
Recent photoemission data~\cite{Okazaki2004} display the quasi-particle peak at the Fermi level and
high energy structure associated with the lower Hubbard band about -1~eV.
The effective electron mass, $m^*/m$, extracted from optics~\cite{Okazaki2006}
is about 4.3 which is larger then the value
observed for Sr$_x$Ca$_{1-x}$VO$_3$ characterized as strongly correlated paramagnetic
metals~\cite{Pavarini2004Sekiyama2004}.
Recent LDA+DMFT investigations~\cite{Biermann2005Liebsch2005Laad2006aLaad2005}
of vanadium oxide have paid an attention to
the spectral properties of the compound and have showed a good agreement with photoemission
and optical data.

In the present work we have used state of the art LDA+DMFT method~\cite{Anisimov1997Lichtenstein1998}
to study the physics of \vo2 in metallic rutile phase with the special focus
on the magnetic properties.
Temperature dependence of the calculated magnetic susceptibility $\chi(T)$ reproduces
with a good accuracy Curie-Weiss law
$\chi_{CW}(T) = p_{eff}^2/3(T-\Theta)$ and is in an agreement with experimental data.
Calculations results for the spectral function are typical for strongly correlated metal
with well developed Hubbard bands in addition to quasi-particle metallic peak at Fermi level
and they agree well with earlier studies~\cite{Biermann2005Liebsch2005Laad2006aLaad2005}.
Calculated value of effective electron mass enhancement factor  $m^*/m \approx 2$
proves a relatively large strength of correlation effects for metallic phase of \vo2.
Therefore, vanadium dioxide in rutile phase is the strongly correlated metal with local magnetic moments.

At high temperatures, vanadium dioxide crystallizes as rutile
shown on the Fig.~\ref{fig:structure_r}. It has the $P4_2/mnm$ space group and
details of lattice constants and atomic positions can be found in Ref.~\onlinecite{McWhan1974}.
Each vanadium atom is surrounded by a oxygen's octahedron resulting in splitting
of $d$ level to triply degenerate \t2g and doubly degenerate \egsi states.
Additional tetragonal distortion present in the structure leads to further
lifting of degeneracy of \t2g level to \a1g and $\pi$ states ($d_{||}$ and $d_{\pi}$
according to Gooenough's notations~\cite{Goodenough1971}).
In the local coordinate basis shown on the Fig.~\ref{fig:structure_r} by (double) primed
axes for vanadium atoms, \a1g state corresponds to the $d_{x'y'}$ orbital,
$\pi$ states correspond to $d_{x'z'}$ and $d_{y'z'}$ and \egsi to $d_{z'^2}$ and
$d_{x'^2-y'^2}$ orbitals. The local $z'$ axis is chosen to be pointed to
apical oxygen atoms and local $x'$ and $y'$ are chosen to be pointed to
the planar oxygens. This local coordinate system can be obtained by the rotation
to the Euler's angles $(\pi/4,-\pi/2,\pi/4)$.
The double primed local coordinate axes are obtained
by a proper symmetry operation for corresponding vanadium atoms.

\begin{figure}
  \centering
  \includegraphics[clip=true, width=0.4\textwidth]{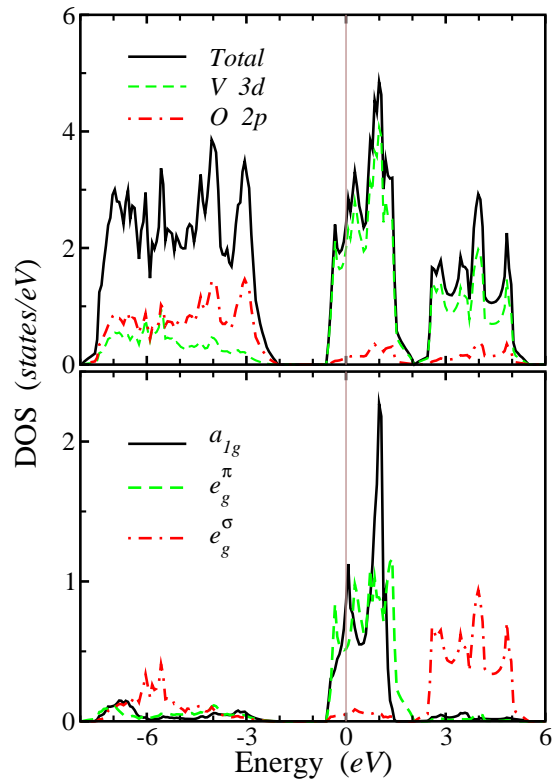}
  \caption{(Color online)  LDA density of states for the rutile phase of \vo2.
           Top panel shows total, V 3$d$ and O 2$p$ DOSes by (black) solid,
           (green) dashed and (red) dot-dashed lines, respectively.
           Bottom panel shows partial DOSes of different symmetry for vanadium atom
           (see legends for notations)}
  \label{fig:lda_dos_r}
\end{figure}

We used TB-LMTO method~\cite{Andersen1975Andersen1984Andersen1986}
to calculate electronic structure of \vo2.
The results obtained are presented on the Fig.~\ref{fig:lda_dos_r} and
have a good agreement with earlier studies~\cite{Eyert2002}.
Total and partial (per atom) density of states are shown on the top panel
(for details of color-coding see caption).
One can clearly see that the states crossing Fermi level are of vanadium \t2g
symmetry mainly and spread from -0.5~eV to 2~eV.
They are well separated by a gap of 1.4~eV from occupied bonding states
of mixed O 2$p$ and V \egsi character.
The anti-bonding combination is located above 2~eV and is also separated by a tiny gap
from the \t2g bands. The orbitally (symmetry) decomposed partial DOSes
for vanadium atom in the local coordinate system mentioned above
are presented on the bottom panel of Fig.~\ref{fig:lda_dos_r}.
\a1g state is more narrow than $\pi$ one and has a one-dimensional-like shape.

We constructed low-energy Hamiltonian for V \t2g states using the $N$MTO method~\cite{Andersen2000}.
It has a dimension of 6 and will be used in LDA+DMFT calculation.
The eigenvalues of the $N$MTO Hamiltonian coincide with the LMTO's one by construction
in the energy interval close to the Fermi level and shape of the density of states remains the same
(not presented).
Hopping integrals for the Wannier function of \a1g, $\pi_1$ and $\pi_2$ 
($d_{x'y'}, d_{x'z'}$ and $d_{y'z'}$ in local notations) symmetries
are presented in the Table~\ref{tab:R_hoppings} for the vanadium atoms in chain
(along $c_R$ axis) and between chains (upper and lower parts of table, respectively).
One can immediately see that \a1g electron can easily jump to the same orbital
on the nearest neighbor atom and it is the largest (intra chain) hopping.
The probability to hop to another chains is much smaller for \a1g electron.
Overall structure of \a1g hoppings assumes one-dimensional-like physics
where the largest transfer integral defines energy splitting between peaks and
the rest is responsible for asymmetry of density of states and its broadening
(see \a1g partial DOS on the bottom panel of Fig.~\ref{fig:lda_dos_r}).
The structure of the $\pi$ hoppings is more uniform in space that
produces more three dimensional shape of density of states.

\begin{table}[t]
  \vspace{5mm}
  \begin{center}
    \begin{tabular}{|c|c||c|r|r|r|}
      \hline
            & $d$, \AA &     & \a1g & $\pi_1$ & $\pi_2$   \\
      \hline
      \hline
   Intra    &      &  \a1g   &  310 &    0    &    0     \\
   chain    & 2.85 & $\pi_1$ &   0  &    63   &  155     \\
   hopping  &      & $\pi_2$ &   0  &   155   &   63     \\
      \hline
      \hline
   Inter    &      &  \a1g   &   6  &    0    &    0     \\
   chain    & 3.52 & $\pi_1$ &  40  &   177   &   26     \\
   hopping  &      & $\pi_2$ &  40  &   177   &   26     \\
      \hline
    \end{tabular}
  \end{center}
  \caption{Hopping integrals for the rutile phase (in meV).
           The distance between vanadium atoms is in second column.
           The largest hoppings to another distances are smaller than 25~meV and not shown here.}
  \label{tab:R_hoppings}
\end{table}

To take into account electron-electron correlation in the partially
filled $d$-band we use LDA+DMFT method~\cite{Anisimov1997Lichtenstein1998}.
It combines the material specific aspect in LDA and accurate treatment
of local Coulomb intraction in DMFT~\cite{Kotliar2006,Georges1996}. The later is achieved via mapping
of the lattice problem with many degrees of freedom to the quantum impurity
embedded to the time dependent self-consistent bath (for details about LDA+DMFT method and
implementations see e.g. Refs~\onlinecite{Lechermann2006Anisimov2005}).
For the solution of the quantum impurity problem the modern multi-orbital Hirsch-Fye
quantum Monte-Carlo method has been used~\cite{Poteryaev2007}.
In our LDA+DMFT calculations we use the value of screened Coulomb interaction, $U$=4~eV,
and the value of Hund's exchange, $J$=0.68~eV. In the quantum Monte Carlo
calculations we keep $\Delta\tau\equiv\beta/L$=0.25 for all temperatures and
of order $10^6 \div 10^7$ QMC steps to satisfy ergodicity.

\begin{figure}
  \centering
  \includegraphics[clip=true, width=0.4\textwidth]{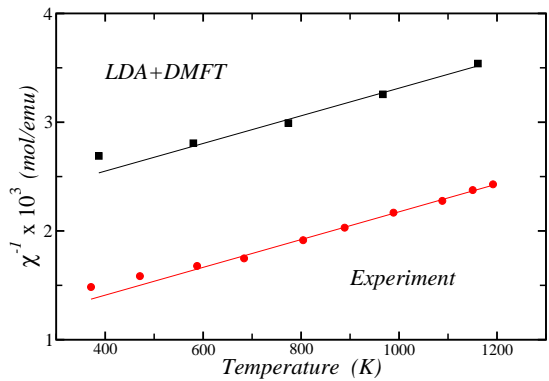}
  \caption{(Color online)  Inverse of uniform magnetic susceptibility.
           LDA+DMFT results are shown by (black) squares, the experimental
           data by Zylbersztejn {\it et al.}~\cite{Zylbersztejn1975} are (red) circles.
           Lines are result of linear fit of data above 500~K.}
  \label{fig:chi_r}
\end{figure}

We start the presentation of the LDA+DMFT results from magnetic properties.
Following the linear response idea, the uniform magnetic susceptibility can be calculated
directly by adding to the Hamiltonian an external magnetic field, $\mu_B H_z$,
and measuring magnetization of compound. Then, the uniform magnetic susceptibility is
\begin{equation}
  \chi(T) = \frac{m(T)}{H_z},
\end{equation}
where $m(T) = \sum_m n_m^{\uparrow} - n_m^{\downarrow}$ is a magnetization at given temperature.
The few magnetic fields of order $\sim0.02\div0.1$~eV were used to check and satisfy
the condition of linearity of magnetization.
The inverse of calculated uniform magnetic susceptibility and experimental data
extracted from Ref.~\onlinecite{Zylbersztejn1975} are shown
on the Fig.~\ref{fig:chi_r}. One can clearly see that these inverse quantities agree
accurately to the shift along ordinate axis. This implies that the calculated and
experimental effective magnetic moments are in good agreement.
The theoretical estimate obtained by the fit to the Curie-Weiss law
\begin{equation}
  \chi_{CW}(T) = \frac{p_{eff}^2}{3(T-\Theta)}
\end{equation}
gives the value of the effective magnetic moment, $p_{eff}^{theor} = 1.54 \mu_B$,
in a good agreement with the experimental value  $p_{eff}^{exp} = 1.53 \mu_B$
that is close to ideal value for V$^{4+}$ ion with the spin $S=1/2$,
$p_{eff} = 1.73 \mu_B$.
To make the fit shown by lines we have used the data points above 500~K where the linearity
of $\chi^{-1}$ is more pronounced for theoretical and experimental data as well.
The use of complete data set leads to the slightly higher values of magnetic moments~\cite{fullT}.
The theoretical Curie-Weiss temperature $\Theta$ is almost twice larger then its experimental 
counterpart (-1600~K and -700~K correspondingly), that is connected to the mean-field
nature of the calculations and, hence, absence of the collective excitations.

\begin{figure}
  \centering
  \includegraphics[clip=true, width=0.4\textwidth]{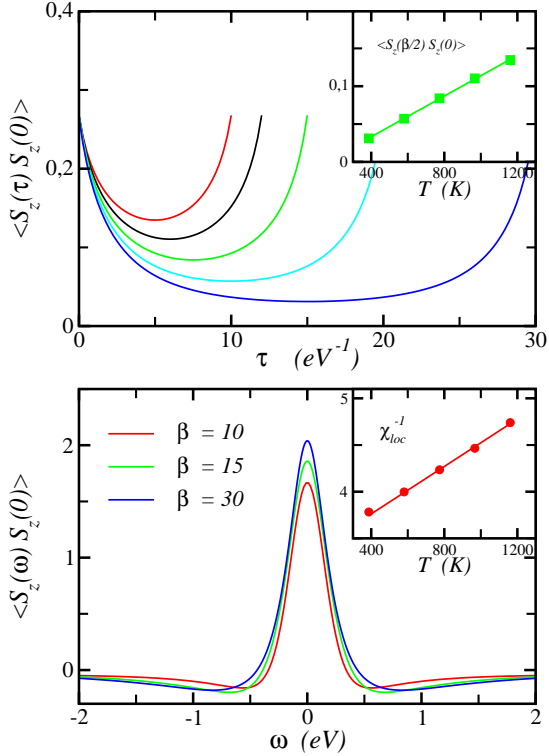}
  \caption{(Color online)  The local spin-spin correlation function for different temperatures
           in imaginary time domain (upper panel) and its analytical continuation
           to the real energies (lower panel). Upper inset shows the dependence of
           this correlator at $\beta/2$ with temperature. Lower inset is an inverse of the local
           susceptibility (for details see text).}
  \label{fig:chi_local_r}
\end{figure}

We have also calculated a local spin-spin correlation function,
$\langle S_z(\tau)S_z(0) \rangle$ at different temperatures shown on the Fig.~\ref{fig:chi_local_r}.
Albeit this is a pure local quantity and cannot be directly compared to the experimental data,
it gives a sound source of information about magnetic properties of the system.
At low temperatures, in the Fermi liquid regime, the local spin-spin correlation function
behaves as $[T/sin(\pi\tau T)]^2$ at intermediate times, $\tau$, while
it saturates to the constant value for the insulator or metal with a very well developed
local magnetic moment~\cite{Georges1996,Werner2008}.
At middle point, the correlator is
$\langle S_z(\beta/2)S_z(0) \rangle \approx T^{\alpha}$, with exponent $\alpha$=2
for Fermi liquid and zero for the insulator.
On the upper panel of the Fig.~\ref{fig:chi_local_r}, from the time dependency
one can see that the correlation function is not saturated.
At the same time, the $\langle S_z(\beta/2)S_z(0) \rangle$ quantity plotted
in the inset of upper panel shows the linear behavior with the exponent $\alpha$=1 different
from both insulating and metallic regimes. This observation suggests
that the formation of the local magnetic moment in the metallic phase
is not completely finished~\cite{Werner2008}.
The lower panel of the Fig.~\ref{fig:chi_local_r} shows the analytical continuation
of the local spin-spin correlation function to the real energy axis~\cite{analcont}.
The width of the peak at zero frequency is inverse proportional to a ''lifetime``
of the local magnetic moment and it is much larger than that of the $\alpha$-iron
in paramagnetic phase~\cite{Katanin2010} which is closer to the localized limit.
The inset in the lower panel shows the inverse of the local magnetic susceptibility,
$\chi_{loc}(T) = \int_0^{\beta} d \tau \langle S_z(\tau)S_z(0) \rangle$
versus temperature. This quantity has a noticeable linear behavior and by fitting it
to the Curie law one obtains
the value of the effective local magnetic moment, $p_{loc}=1.54 \mu_B$ and
critical temperature.
The comparison of the local and uniform susceptibilities gives a rough estimation
of the degree of spacial correlations.
The close values of the extracted effective magnetic moments, $p_{eff}^{theor}$ and $p_{loc}$,
assumes that the vanadium dioxide at high temperatures has a localized type of magnetism.

\begin{figure}
  \centering
  \includegraphics[clip=true, width=0.4\textwidth]{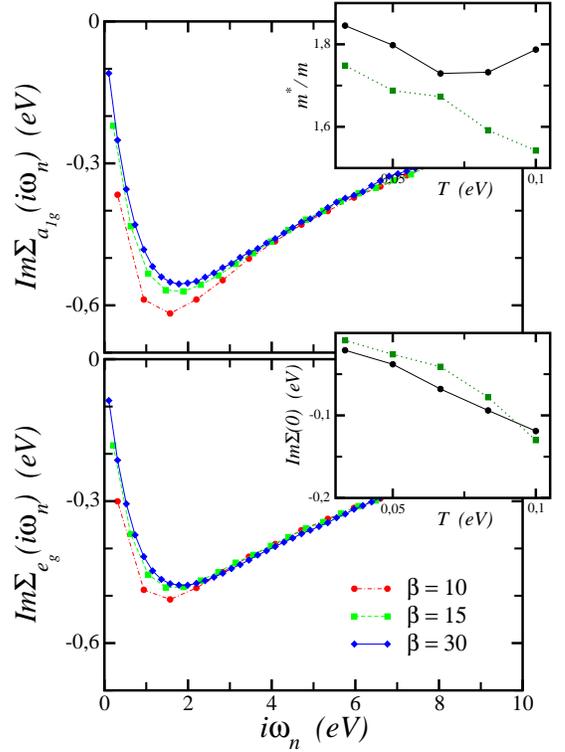}
  \caption{(Color online) Imaginary part of self-energy for \a1g (top) and 
           \eg (bottom) states in metallic phase.
           Solid (blue), dashed (green) and dot-dashed (red) lines present
           $\beta$=30, 15 and 10~eV$^{-1}$ respectively ($\sim$390, 770 and 1160~K).
           Upper inset shows the effective mass, $m^*/m$, for \a1g and \eg
           orbitals by (black) circles and (dark green) squares, respectively.
           Lower insets shows the imaginary part of self-energy
           extracted to zero energy, $\Im\Sigma(0)$.
           Color-coding is the same as in previous inset.}
  \label{fig:sigma_r}
\end{figure}

Now, we would like to turn our discussion to single particle properties.
The imaginary part of the self-energy for \a1g and $\pi$ orbitals are shown
on the Fig.~\ref{fig:sigma_r}. The behavior of the imaginary part for both orbitals
is of Fermi-liquid type. At small real frequencies, in Fermi-liquid regime,
the imaginary part is
$\Im\Sigma(\omega) = -\Gamma T^2 - B \omega^2 $ that transforms to the linear behavior
on the Matsubara axis, $\Im\Sigma(i\omega_n) = - \Gamma T^2 - B \omega_n$,
observable on the figure. Additionally, the mass enhancement factor
\begin{equation}
  m^* / m = 1 - \frac{\Im\Sigma(i\omega_0)}{\omega_0}
\end{equation}
as a function of temperature shows a saturation (upper inset of Fig.~\ref{fig:sigma_r})
indicating the Fermi-liquid behavior~\cite{Jarrell1993} of moderately correlated metal
with $m^*/m \approx$ 1.8 and 1.7 for \a1g and $\pi$ orbitals, respectively.
The lower inset shows the extracted to zero imaginary part of self-energy, $\Im\Sigma(0)$
versus temperature. One can see that it goes quadratically to zero at low temperatures
once again confirming Fermi-liquid nature of compound.
From these data one can conclude that the coherence temperature is about 700~K for both orbitals.

\begin{figure}[*tbh]
  \centering
  \includegraphics[clip=true, width=0.4\textwidth]{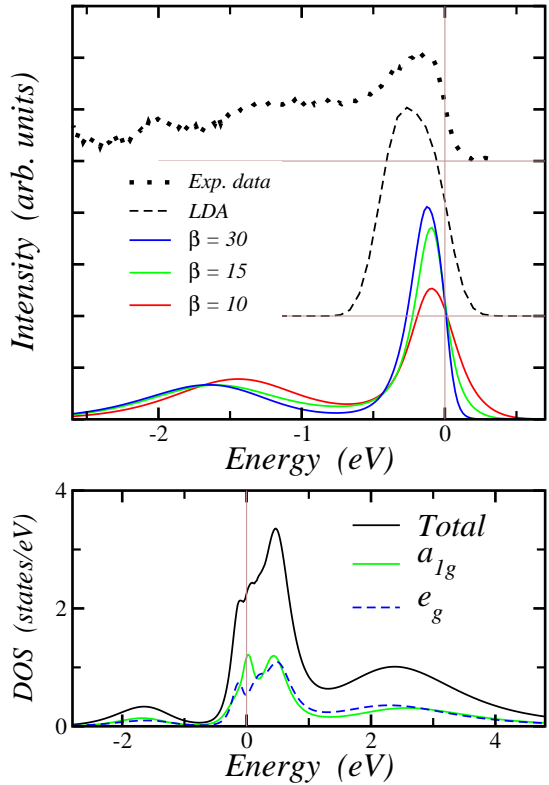}
  \caption{(Color online)  Top panel shows comparison of the photoemission data
  	   extracted from the Ref.~\onlinecite{Okazaki2004} and calculated
  	   LDA and LDA+DMFT results for various temperatures.
  	   Bottom panel is total and orbitally resolved LDA+DMFT spectral functions
  	   for $\beta$=30~eV$^{-1}$ (T$\approx$390~K).}
  \label{fig:pes_r}
\end{figure}

The comparison of the photoemission data~\cite{Okazaki2004} (PES)
with the calculated LDA and LDA+DMFT results are presented on the upper panel of the Fig.~\ref{fig:pes_r}.
One can clearly observe a good agreement of the LDA+DMFT (color lines) and
PES data (dots).
There is a quasi-particle peak at the Fermi level and lower Hubbard band at higher energies.
The later is an additional evidence of the correlated nature of the metallic rutile phase.
Temperature dependence of the LDA+DMFT spectra shown by different colors is also
in good qualitative agreement with the experiment (see Fig.~3 of the Ref.~\onlinecite{Okazaki2004}):
the spectral weight is transferred from quasi-particle peak to higher energies and
becomes smaller with decrease of temperature.
At the same time, the LDA results shown by the dashed line describe
the quasi-particle feature at the Fermi level
with a large overestimation of the peak weight and complete absence of the high energy incoherent hump.
LDA+DMFT spectral function for lowest calculated temperature, $T \approx$~390 K is presented
on the lower panel of the Fig.~\ref{fig:pes_r}.
One can clearly see the three peaks structure which is typical for correlated materials.
The quasi-particle peak lies at/above the Fermi level and low and
upper Hubbard bands are located at higher frequencies, at -1.5~eV and $\sim$2.3~eV.
The width of the quasi-particle peak is reduced to $\sim$1.1~eV comparing with 2.1~eV value
for non-interacting band width and is almost independent
with temperature which is compatible with the weak temperature dependence of
the quasi-particle weight, $Z = m/m^*$, for both orbitals (see upper inset of Fig.~\ref{fig:sigma_r}).
One should note here that the spectral functions for higher temperatures are similar
to described above. The difference is that the shape of the quasi-particle peak becomes more complicated
due to the decrease (in absolute value) of the imaginary part of self-energy at zero energy
(see lower inset of the Fig.~\ref{fig:sigma_r}),
and hence, displaying the features of original LDA spectral function smoothed by high temperature
at $\beta$=10~eV$^{-1}$.

In conclusion we have studied the influence of Coulomb correlation effects on magnetic and
spectral properties of \vo2 in metallic rutile phase by LDA+DMFT method.
The calculation results show typical strongly correlated metal behavior for self-energy
with a sizable spectral weight transfer to Hubbard bands and electronic states renormalization
near the Fermi level with effective mass $m^*/m \approx$ 2. 
The uniform and local magnetic susceptibility calculation obey Curie-Weiss temperature dependence
with effective magnetic moment value close to ideal ionic value corresponding to $d^1$ configuration.
A good agreement of calculated and measured spectral and magnetic properties allows to conclude that
\vo2 in rutile phase is in strongly correlated metal state
with local magnetic moments formed by $d$-electrons.

The authors thank A. Georges and S. Biermann who were at the initial stage of this work
and A. Katanin and A. Lichtenstein for useful discussions.
AP thanks to the Marie Curie grant MIF1-CT-2006-021820.
This work was supported by the Russian Foundation for
Basic Research (Projects Nos. 10-02-00046a and 09-02-00431a),
the Dynasty Foundation, the fund of the President of the Russian Federation
for the support of scientific schools NSH 1941.2008.2, the Program of
the Russian Academy of Science Presidium ``Quantum microphysics of
condensed matter'' N7, Russian Federal Agency for Science and Innovations
(Program ``Scientific and Scientific-Pedagogical Trained of the Innovating
Russia'' for 2009-2010 years), grant No. 02.740.11.0217, MK-3758.2010.2.


\begin{thebibliography}{99}
  \bibitem{Imada1998}  M. Imada, A. Fujimori, and Y. Tokura,
    Rev. Mod. Phys. \textbf{70}, 1039 (1998).
    
  \bibitem{Villeneuve1977Pouget1974}  J.P. Pouget \emph{et al.}, Phys. Rev. B \textbf{10}, 1801 (1974);
      G. Villeneuve \emph{et al.}, Journal of Physics C: Solid State Physics \textbf{10}, 3621 (1977).
      
  \bibitem{Chain1991}  E.E. Chain, Appl. Opt. \textbf{30}, 2782 (1991).
  
  \bibitem{Berglund1969}  C.N. Berglund and H.J. Guggenheim, Phys. Rev. \textbf{185}, 1022 (1969).
    
  \bibitem{Goodenough1971}  J. B. Goodenough, Journal of Solid State Chemistry \textbf{3}, 490 (1971).
  
  \bibitem{Rice1994Wentzcovitch1994}  T.M. Rice, H. Launois, and J.P. Pouget,
    Phys. Rev. Lett. \textbf{73}, 3042 (1994);
      R.M. Wentzcovitch, W.W. Schulz, and P.B. Allen, Phys. Rev. Lett. \textbf{72}, 3389 (1994).
      
  \bibitem{Eyert2002}  V. Eyert, Annalen der Physik \textbf{11}, 650 (2002).
  
  \bibitem{Okazaki2004}  K. Okazaki \emph{et al.}, Phys. Rev. B \textbf{69}, 165104 (2004).
  
  \bibitem{Okazaki2006}  K. Okazaki \emph{et al.}, Phys. Rev. B \textbf{73}, 165116 (2006).
  
  \bibitem{Pavarini2004Sekiyama2004}  E. Pavarini \emph{et al.},
    Phys. Rev. Lett. \textbf{92}, 176403 (2004);
      A. Sekiyama \emph{et al.}, Phys. Rev. Lett. \textbf{93}, 156402 (2004).
      
  \bibitem{Biermann2005Liebsch2005Laad2006aLaad2005}  S. Biermann \emph{et al.},
    Phys. Rev. Lett. \textbf{94}, 026404 (2005);
      A. Liebsch, H. Ishida, and G. Bihlmayer, Phys. Rev. B \textbf{71}, 085109 (2005);
      M.S. Laad, L. Craco, and E. M\"uller-Hartmann, Phys. Rev. B \textbf{73}, 195120 (2006);
      M.S. Laad, L. Craco, and E. M\"uller-Hartmann, Europhysics Letters \textbf{69}, 984 (2005).
  
  \bibitem{Anisimov1997Lichtenstein1998}  V.I. Anisimov \emph{et al.},
    Journal of Physics: Condensed Matter \textbf{9}, 7359 (1997);
      A.I. Lichtenstein and M.I. Katsnelson, Phys. Rev. B \textbf{57}, 6884 (1998).
      
  \bibitem{McWhan1974}  D.B. McWhan \emph{et al.}, Phys. Rev. B \textbf{10}, 490 (1974).
  
  \bibitem{Andersen1975Andersen1984Andersen1986}  O.K. Andersen, 
    Phys. Rev. B \textbf{12}, 3060 (1975);
      O.K. Andersen, O. Jepsen and M. Sob, The Electronic Band Structure and Its Applications,
       \emph{Springer-Verlag}, Berlin, 1986.
       
  \bibitem{Andersen2000}  O.K. Andersen and T. Saha-Dasgupta, Phys. Rev. B \textbf{63}, R16219 (2000).
  
  \bibitem{Kotliar2006} W. Metzner and D. Vollhardt, Phys. Rev. Lett. \textbf{62}, 324 (1989); 
     G. Kotliar and D. Vollhardt, Phys. Today \textbf{57}, 53 (2004); 
     G. Kotliar \emph{et al.}, Rev. Mod. Phys. \textbf{78}, 865 (2006).
  
  \bibitem{Georges1996} A. Georges \emph{et al.}, Rev. Mod. Phys. \textbf{68}, 13 (1996).
  
  \bibitem{Lechermann2006Anisimov2005}  F. Lechermann \emph{et al.},
    Phys. Rev. B \textbf{74}, 125120 (2006);
      V.I. Anisimov \emph{et al.}, Phys. Rev. B \textbf{71}, 125119 (2005).

  \bibitem{Poteryaev2007}  A.I. Poteryaev \emph{et al.}, Phys. Rev. B \textbf{76}, 085127 (2007).
  
  \bibitem{Zylbersztejn1975}  A. Zylbersztejn and N.F. Mott, Phys. Rev. B \textbf{11}, 4383 (1975).
  
  \bibitem{fullT} The use of the complete data set in the fitting procedure leads to
     a larger values for the effective magnetic moment and transition temperature
     for experimental and theoretical results:
     $p_{eff}^{theor} = 1.64 \mu_B$, $\Theta^{theor} =$ -1980 K and
     $p_{eff}^{exp} = 1.6 \mu_B$, $\Theta^{exp} =$ -860 K.
  
  \bibitem{Werner2008}  P. Werner \emph{et al.}, Phys. Rev. Lett. \textbf{101}, 166405 (2008).
  
  \bibitem{analcont}  H.J. Vidberg and J.W. Serene, Journal of Low Temperature Physics \textbf{29}, 179 (1977).
  
  \bibitem{Katanin2010}  A.A. Katanin \emph{et al.}, Phys. Rev. B \textbf{81}, 045117 (2010).
  
  \bibitem{Jarrell1993}  M. Jarrell and T. Pruschke, 
    Zeitschrift f\"ur Physik B Condensed Matter \textbf{90}, 187 (1993).
  
\end{thebibliography}

\end{document}